\documentclass[11pt,letterpaper]{article}
\usepackage{epsfig,amssymb,amsmath, multirow, url, tcolorbox,booktabs, enumitem}

\usepackage{hyperref}
\usepackage{graphicx}
\usepackage{enumitem}
\usepackage{natbib}
\usepackage[utf8]{inputenc}
\usepackage{authblk}
\usepackage{lscape}
\usepackage{adjustbox}
\newsavebox{\theorembox}
\newsavebox{\lemmabox}
\newsavebox{\claimbox}
\newsavebox{\factbox}
\newsavebox{\corollarybox}
\newsavebox{\examplebox}
\newsavebox{\remarkbox}
\newsavebox{\assbox}
\newsavebox{\propositionbox}
\newsavebox{\problembox}
\newsavebox{\defbox}
 
\savebox{\theorembox}{\noindent\bf Theorem}
\savebox{\lemmabox}{\noindent\bf Lemma}
\savebox{\factbox}{\noindent\bf Fact}
\savebox{\corollarybox}{\noindent\bf Corollary}
\savebox{\examplebox}{\noindent\bf Example}
\savebox{\assbox}{\noindent\bf Assumption}
\savebox{\propositionbox}{\noindent\bf Proposition}
\savebox{\problembox}{\noindent\bf Problem}
\savebox{\defbox}{\noindent\bf Definition}

\newtheorem{@assumption}{\bf Assumption}[section]

 \newtheorem{@remark}{\bf Remark}[section]
 \newenvironment{remark}{\begin{@remark}\rm}{\end{@remark}}

\newcommand{\argmin}{\mathop{\rm argmin}}

\def\blot{\quad {$\vcenter{\vbox{\hrule height.4pt
             \hbox{\vrule width.4pt height.9ex \kern.9ex \vrule
width.4pt}
             \hrule height.4pt}}$}}

\usepackage{multirow}
\providecommand{\keywords}[1]
{
{  \small	
  {\textit{Keywords---}} #1}
}
\usepackage{cleveref}
\crefname{equation}{equation}{equations}
\Crefname{equation}{Equation}{Equations}% For beginning \Cref
\crefrangelabelformat{equation}{(#3#1#4--#5#2#6)}
\crefmultiformat{equation}{equations (#2#1#3}{, #2#1#3)}{#2#1#3}{#2#1#3}
\Crefmultiformat{equation}{Equations (#2#1#3}{, #2#1#3)}{#2#1#3}{#2#1#3}
\usepackage{caption}
\usepackage{subcaption}
\usepackage{graphicx}
\usepackage{rotating}
\usepackage{tikz}

\begin{document}
\title{A co-segmentation algorithm to predict emotional stress from passively sensed mHealth data}

\author[1,2]{Younghoon Kim}
\author[1]{Sumanta Basu}
\author[2]{Samprit Banerjee$^\dagger$}

\affil[1]{Cornell University}
\affil[2]{Weill Cornell Medicine}

\date{\today}
\maketitle
\def\thefootnote{$\dagger$}\footnotetext{Corresponding author. Email: sab2028@med.cornell.edu}

\begin{abstract}
We develop a data-driven co-segmentation algorithm of passively sensed and self-reported active variables collected through \textcolor{black}{smartphones to identify emotionally stressful states in middle-aged and older patients with mood disorders undergoing therapy, some of whom also have chronic pain}. Our method leverages the association between the different types of time series. These data are typically non-stationary, with meaningful associations often \textcolor{black}{occurring} only over short time windows. Traditional machine learning (ML) methods, \textcolor{black}{when} applied globally on the entire time series, often \textcolor{black}{fail to capture these time-varying local patterns}. Our \textcolor{black}{approach} first segments the passive sensing variables by detecting their change points, then \textcolor{black}{examines} segment-specific associations with the active variable to identify co-segmented periods that exhibit distinct \textcolor{black}{relationships} between stress and passively sensed measures. \textcolor{black}{We then} use these periods to predict future emotional stress states \textcolor{black}{using} standard ML methods. By \textcolor{black}{shifting the unit of analysis} from individual time points \textcolor{black}{to} data-driven segments of time and allowing for different associations in different segments, our algorithm helps detect patterns that only exist \textcolor{black}{within} short-time windows. We \textcolor{black}{apply} our method to detect periods of stress in \textcolor{black}{patient data} collected \textcolor{black}{during} ALACRITY Phase I study. \textcolor{black}{Our findings indicate} that \textcolor{black}{the} data-driven segmentation algorithm \textcolor{black}{identifies} stress periods more accurately than traditional ML methods that do not \textcolor{black}{incorporate} segmentation.
\end{abstract}

\keywords{Stress detection, mental health, mHealth, change point detection, machine learning, classification}

%%%%%%%%%%%%%%%%%%%%%%%%%%%%%%%%%%%%%%%%%%%%%%%%%%%%%%%%%%%%%%%%%%%

\section{Introduction}
\label{se:intro}

As we enter the ubiquitous era of mobile health (mHealth) devices, such as wearables and smartphones, the potential use of passively sensed data from these devices is becoming increasingly relevant in health studies. \textcolor{black}{Passively sensed data refer to information recorded by mHealth device sensors (e.g., accelerometer, GPS, microphone, etc.) without active input from the user \citep{mendes2022sensing,moura2023digital}.}

In particular, passively sensed data on physical activity, sociability, sleep patterns \textcolor{black}{hold} great potential for studying behavioral and mental health. The primary reason is that individualized measures of behavioral patterns can help identify mental states, \textcolor{black}{aiding in diagnosis and treatment} related disorders. For example, identifying emotional stress states and discovering digital behavioral markers \textcolor{black}{associated with these} emotional states, such as psychomotor activities \citep{sobin1997psychomotor}, attention-distractibility \citep{forster2016establishing}, and moods \citep{clark1988mood}, is \textcolor{black}{a critical challenge} in mental health research. \textcolor{black}{These} digital markers can indicate the onset of serious mental health conditions, such as depression and anxiety \citep{vahia2013diagnostic}, and \textcolor{black}{enable the development of} digital interventions that can be delivered to the right patient and \textcolor{black}{at} the right time.  \citep{barkley2014attention}. In this paper, we focus on \textcolor{black}{detecting} periods of emotional stress using passively sensed mHealth data.

Emotional stress is a \textcolor{black}{state of} psychological tension and discomfort caused by various individual factors, such as internal conflicts, frustrations, diminished self-esteem, grief, and environmental factors, including danger, threats, and the loss of security \citep{vandenbos2007apa}. In \textcolor{black}{this study, we focus on middle-aged and older patients. All of them have mood disorders and a subset of them also have chronic pain.} For \textcolor{black}{these individuals, prolonged} exposure to emotional stress periods is known to increase the risk of both mental and physical symptoms, including depression, anhedonia, dementia, and actual pains \citep{murray1996global,bair2003depression,arnow2006comorbid}. Once they \textcolor{black}{enter these stressful periods}, negative mood and observable symptoms also tend to be \textcolor{black}{persist over time} \citep{mazure1998life,kendler2003life,monroe2005life}. Thus, preventing \textcolor{black}{patients} from \textcolor{black}{entering} stressful periods is a critical \textcolor{black}{goal} of psychotherapeutic interventions \citep{hammen2005stress,fiske2009depression}.

\textcolor{black}{Studies conducted to detect periods of emotional stress typically share several key characteristics.} First, supervised laboratory settings are often used to detect individuals' \textcolor{black}{stress periods}. In these cases, experimenters \textcolor{black}{introduce} artificial stressors, such as cognitive loads or challenging tasks \citep{healey2005detecting,setz2009discriminating}, \textcolor{black}{which induces short-term stress}. Second, \textcolor{black}{these studies primarily focus on young and healthy participants} \citep{ben2015next,sagl2019wearables,sheikh2021wearable}. Third, \textcolor{black}{emotional stress is typically detected using} specialized sensors. For example, hip-worn accelerometers or gyroscopes are used to measure movement vectors or acceleration \citep{leroux2021wearable,hickey2021smart}. Physiological measures, such as galvanic skin responses (GSR), electrocardiograms, or respiratory rates, \textcolor{black}{require appropriate monitoring devices} \citep{iqbal2022stress,ritsert2022heart}. Lastly, \textcolor{black}{the identification of emotional stress periods often relies on biologically informed rules pre-defined by clinical experts} \citep{kyriakou2019detecting,mansi2021measuring}.

Compared to existing studies on stress detection, \textcolor{black}{identifying} emotional stress in patients with \textcolor{black}{mood disorders} requires long-term observations in the patient's natural environment \textcolor{black}{to pinpoint} the right moments of clinical interventions \citep{zis2017depression,maresova2019consequences}. To this end, using passively sensed mHealth data collected through commercially available mHealth devices shows promise. Additionally, to establish a \textcolor{black}{supervised-like} setting in daily life without \textcolor{black}{needing} experimenter observation, \textcolor{black}{active variables,} consisting of daily self-reporting items regarding stress levels and related positive and negative \textcolor{black}{emotions}, can be employed. Ecological momentary assessment (EMA) \textcolor{black}{is} a form of active sensing \citep{armey2015ecological,weber2022physiological} that can collect self-reported emotional stress on a daily basis. Therefore, to detect periods of emotional stress, one can leverage the association between active (self-reported stress) and passive variables in a data-driven framework, combined with machine learning (ML) algorithms \citep{mohr2017personal,graham2019artificial}.

\begin{figure}[t]
\centering
\includegraphics[width=0.5\columnwidth,height=0.45\textwidth]{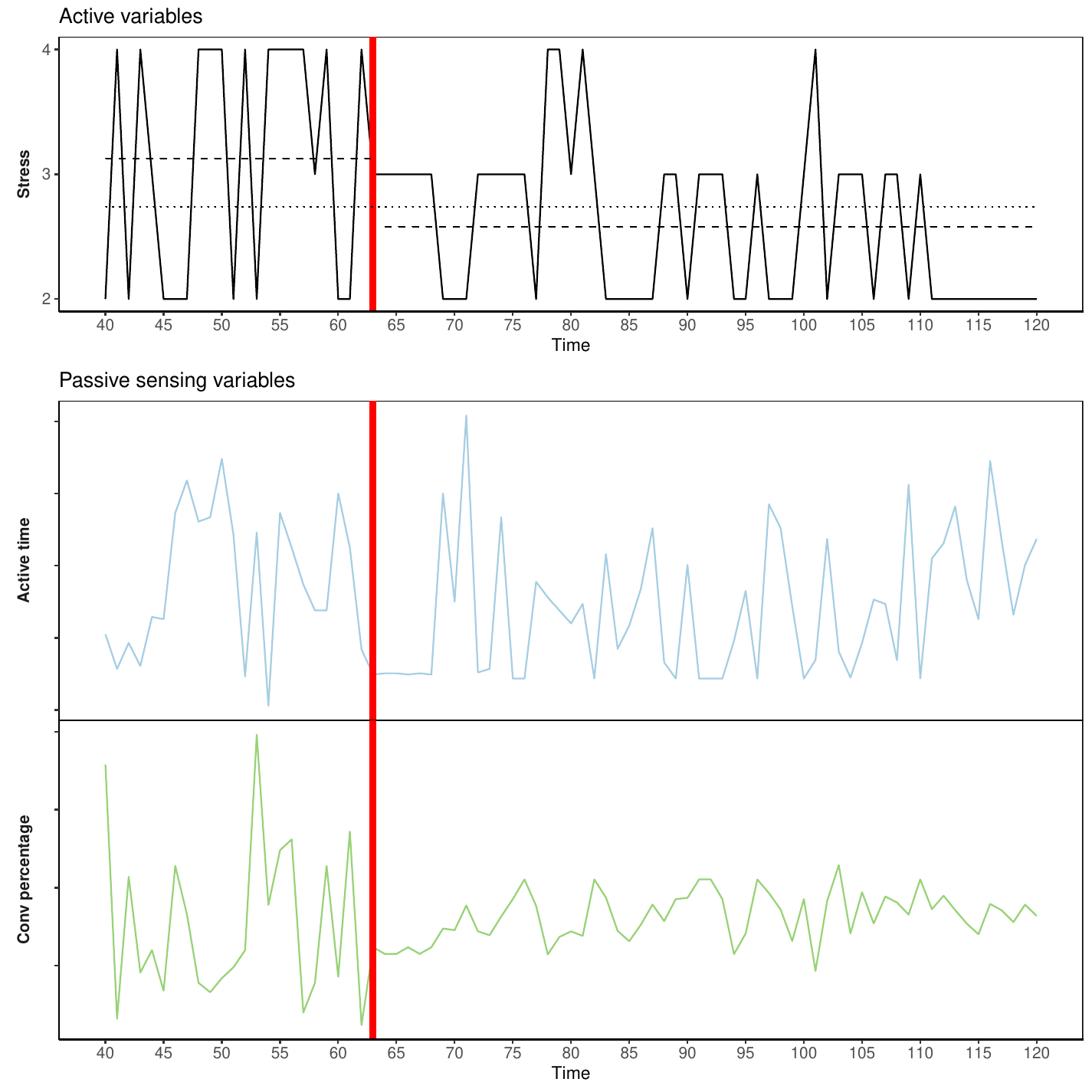}
\caption{Illustration of the portion of \textcolor{black}{daily} sample paths of the stress level $Y$ (top), and two selected passive sensing variables $X$ (bottom). The red vertical line is the manually drawn time point where the significant changes in both types of series are seemingly observed. The black dotted line is the mean of the active variable for all samples in the period. The black dashed lines are the means of the active variable for the samples before and after the time point for the red line.}
\label{fig:seg_nonseg}
\end{figure}

However, merely investigating the association between the variables is not sufficient. The challenge in identifying such a relationship arises from the non-stationary nature of the association. Figure \ref{fig:seg_nonseg} provides an example of sample paths \textcolor{black}{for a} discrete-valued active variable, the stress level in our context, and two continuous-valued passively sensed mHealth data. The association between those two types of variables throughout the entire sample path is barely recognizable. However, the data before and after the time point where the manually drawn red vertical line passes in Figure \ref{fig:seg_nonseg} demonstrates simultaneous changes \textcolor{black}{in the} distributions \textcolor{black}{of} those two variable \textcolor{black}{types}. As illustrated in the figure, the changes in association emerge only when \textcolor{black}{the data are} scrutinized \textcolor{black}{within a} segmented period. Therefore, our study \textcolor{black}{proposes} a data-driven approach to detect the changes in association between two different datasets without \textcolor{black}{manual} inspection. 

\textcolor{black}{The proposed algorithm in this study divides the observations of passive sensing and active variables into several segments, where the segments are determined by identifying time points where the association between active and passive variables changes. In what follows, we use the term co-segments to refer to the segments that are deemed by our algorithm to have the same association.}

\subsection{Related Works}
\label{sse:related}

To our knowledge, there is no similar work that predicts or identifies emotionally stressful periods \textcolor{black}{in patients based on} daily-summarized activity levels. However, since our work \textcolor{black}{focuses on identifying} significant changes in patterns and associating them with the changes in emotional stress states, \textcolor{black}{along with the use of} ML methods to predict the emotional stress states, we can highlight the two relevant groups of literature in terms of \textcolor{black}{methodology}.

\textcolor{black}{The change point detection (CPD) algorithm \citep{truong2020selective,cho2022two}, which is a technique to estimate a set of time points where changes in distributions happen, is crucial in the proposed algorithm. The human activity recognition studies (HAR) \citep{straczkiewicz2021systematic} are similar to our study in that they find changes in patterns of passively sensed data through smartphones, wearable sensors, and installed cameras. However, they primarily focus on tracking and analyzing the lifestyle of the observed individuals rather than exploring individuals' mental health. Furthermore, their algorithms are adapted to their problems since the data used in the study is quite different from ours. For instance, \cite{sadri2017information} developed Information
Gain-based Temporal Segmentation (IGTS) based on Shannon's entropy measures, where the probabilities are derived from occupation time at specific locations over observation time transmitted from cell tower ID (CID). The time points when the sum of the entropies is maximized are identified as change points. Since location transitions do not occur frequently, and users tend to remain in one location once a transition happens, the time plots exhibit only a few jumps and remain relatively flat with minimal fluctuations. \cite{gharghabi2019domain} created the FLUSS (Fast Low-Cost Unipotent Semantic Segmentation) algorithm, which computes distances between shapelets and uses it to compare counts of repetitive patterns in a sliding window when the observations are in temporal and random orders. Then the change points are determined as time points when their discrepancy is maximized. As noticed in their manuscript, the algorithm is effective for data with dense and repetitive patterns (e.g., RFID signals). Therefore, there is a need to adopt more general change point detection algorithms.}

Recent studies on ML algorithms for identifying emotional stress states are also closely related to our work. For instance, \cite{sukei2021predicting} employed a mixture model and a hidden Markov model to extract features that inform the inference of emotional states from various passively sensed and actively collected data. \cite{saylam2023quantifying} utilized deep learning models to predict behavioral markers for office workers using data from wearable devices. \cite{adler2022machine} assessed whether ML methods could be generalized for combined longitudinal study data to predict mental health symptoms. Although these studies \textcolor{black}{focus} on the prediction of self-reported mental health symptoms using \textcolor{black}{passively sensed} data, there is currently no study \textcolor{black}{aiming} at detecting periods of emotional stress in middle-aged and older patients with \textcolor{black}{mood disorders and chronic pain.}

\subsection{Contribution}
\label{sse:contribution}

We propose a data-driven approach that explains contemporaneous changes in \textcolor{black}{the} patterns of the active variable of interest and passively sensed mHealth data. \textcolor{black}{We} apply the proposed algorithm to detect emotional stress periods. The developed framework is expected to contribute to understanding how changes in human activity induce changes in emotional stress.

The proposed framework is also expected to \textcolor{black}{expand} the scope of using mHealth data \textcolor{black}{to include the detection of contemporaneous changes of} patterns in self-report \textcolor{black}{mental} health status and passively sensed data. The proposed data-driven approach also addresses the challenges \textcolor{black}{faced by} existing rule-based approaches.

\subsection{Outline of Paper}
\label{sse:contributions}

The remainder of this article is organized as follows. The proposed algorithm is introduced in Section \ref{se:methods}. The explanation of the data and the application and validation of the algorithm are illustrated in Section \ref{se:application}. We conclude with a discussion in Section \ref{se:conclusion}.

%%%%%%%%%%%%%%%%%%%%%%%%%%%%%%%%%%%%%%%%%%%%%%%%%%%%%%%%%%%%%%%%%%%
\section{Proposed Methods}
\label{se:methods}

\subsection{Data Description and Pre-Processing}
\label{sse:data}

In this section, we illustrate the procedure and results of the algorithm for detecting emotional stress periods. We use the mHealth data generated from the studies conducted by the Weill Cornell ALACRITY Center (P50MH113838). The participants of these studies (three studies in total) were middle-aged and older adults \textcolor{black}{with mood depression undergoing psychotherapy}. The participants whose subject IDs contains `relief' have also chronic pain. Their passively collected mHealth data were recorded through smartphone applications \textcolor{black}{(see Source column in Table \ref{tab:data}) designed to augment} augmented the psychotherapies.

The passively sensed mHealth data \textcolor{black}{consists} of 11 variables \textcolor{black}{categorized into four conceptual groups}:  (1) mobilization (step counts, activity time, duration); (2) sleep (sleep hours, sleep quality), (3) sociability (time in conversation, duration of conversations during activities), and (4) location (travel distance, time spent traveling). Each variable was measured on a different scale such as counts, \textcolor{black}{often in hundreds or more}, or percentages. \textcolor{black}{Note that the measurements were aggregated on a daily basis, ensuring consistency with the frequency of active variables, as surveys were administered to the participants daily.} Due to the \textcolor{black}{varying scales of the passively sensed variables}, we only focus on their original \textcolor{black}{units without normalization}. As seen in Figure \ref{fig:initialization1} below, \textcolor{black}{the data exhibit irregular fluctuations, with some variables pairs showing simultaneous high spikes and sharply returning to the baselines.} The four active variables \textcolor{black}{includes} stress level, arousal, valence, and adjusted stress level. All variables are \textcolor{black}{recorded daily using} the 5 items Likert scales \citep{likert1932technique}. For further details regarding the data utilized in this study, refer to Table \ref{tab:data} \textcolor{black}{in Appendix \ref{sap:data}.}

An important issue in treating passively sensed mHealth data is downward bias. Specifically, passively collected measures from mobile or wearable devices tend to be underestimated \citep{lee2021use}. Two major factors contribute to this bias: either the device was not used at all (non-use), or the participants were not wearing the devices (non-wear), even though they performed the activity. Since \textcolor{black}{data collection occurs} in real-world environments without direct observation, \textcolor{black}{it is difficult to determine} whether participants are accurately wearing the devices \citep{hicks2019best}. Therefore, the non-use case is confused with the non-wear case. However, the non-wear case can be \textcolor{black}{treated} as missing data, which \textcolor{black}{we aim to recover}. To address this, we use 2SpamH \citep{zhang20242spamh}, an algorithm that helps determine whether the bias is due to a non-wear case. Once \textcolor{black}{these} cases are identified, we regard this downward bias as missingness and impute the missing values by using the R package \texttt{MissForest} \citep{stekhoven2012missforest}.

\subsection{Step 1: Estimate Initial Change Points}
\label{sse:step1}

We delve into the procedure for predicting the emotional stress status of the targeted patients in the future through sequential steps. \textcolor{black}{Throughout the remainder of the paper, we assume} that emotional stress states \textcolor{black}{correspond} to hidden states, which are unobserved but can be indirectly inferred from the observed stress levels in \textcolor{black}{co-segments}. \textcolor{black}{Furthermore,} as long as any active variable of interest can be paired with passive sensing variables, the algorithm can be generalized to predict different hidden emotional states such as depression or fatigue in different contexts \citep{mohr2017personal}.

Let $X_{t}\in\mathbb{R}^p$ denote the passive sensing variables for a patient at time $t$ and $Y_t$ be one of the collected active variables at time \textcolor{black}{$t=1,\ldots,T$, where $T$ is the sample length for the patient.} For stress period detection in our example, the variable is the level of stress, ranging from 1 through 5. \textcolor{black}{Define the series of passive sensing variables $\mathcal{L}:=\{X_{t}\}_{t\in\mathcal{T}}$, where $\mathcal{T}=\{1,\ldots,T\}$ is a collection of time indices of the passively sensed data.} The flowchart of the entire algorithm is outlined in Figure \ref{fig:flowchart} in Appendix \ref{sap:overview}. The algorithm consists of five steps.

In Step 1, we construct non-overlapping segments of the series, denoted by $\mathcal{L} = \uplus_{k=1}^{N_{\circ}} \mathcal{L}_{k}$ and $\mathcal{T} = \uplus_{k=1}^{N_{\circ}} \mathcal{T}_{k}$, \textcolor{black}{where $\mathcal{L}_{k}=\{X_{t}\}_{t\in\mathcal{T}_{k}}$ is the $k$th segment of the series $\mathcal{T}=\{X_t\}_{t=1,\ldots,T}$, $\mathcal{T}_{k}=\{t_{k-1},\ldots,t_{k}\}$ is the subset of the time indices belonging to the $k$th segment, and $N_{\circ}$ represents the number of segments after Step 1}. Here, $t_{0}=1$ and $t_{N_{\circ}} = T$. Our goal at this step is to estimate the initial change points on $\mathcal{L}_k$. Once the estimated initial change points of $\mathcal{L}$ are obtained, \textcolor{black}{we have the co-segments $\{X_t,Y_t\}_{t\in\mathcal{T}_k}$ where the time points segmenting the observations are determined by the result of Step 1}. We proceed with the algorithm if $N_0>1$.

We utilize a CPD algorithm developed by \cite{matteson2014nonparametric}. \textcolor{black}{The details on this algorithm are provided in Appendix \ref{sap:matteson}. The algorithm is available in the R package \texttt{ecp} developed by \cite{james2015ecp}.} Note that the algorithm is \textcolor{black}{non-parametric and model-free so that it is applicable to general sequential observations} in that it only relies on the condition of the finiteness of the second moments of the distributions. Changes in passive sensing variables often entail alterations in mean level or variance, which can be adequately captured by changes in the first or second moments between consecutive segments. \textcolor{black}{Furthermore, as seen in Figure \ref{fig:initialization1} below, it is challenging to specify a proper distribution of the passive sensing variables while it is often required for other CPD algorithms. Therefore, this CPD algorithm that works under milder conditions is preferred.}

\begin{remark}\label{rem:CPD}
\textcolor{black}{To our best knowledge, there are only a few CPD methods for count data \cite[e.g.,][]{kirch2014detection,lee2021recent,diop2022poisson} and no available methods for ordinal variables. In addition, due to the non-stationarity of our time series, we prefer non-parametric and model-free CPD methods for multivariate data. \textcolor{black}{The algorithm developed by \cite{matteson2014nonparametric}} is the well-known method to achieve these goals. There are some other methods \cite[e.g.,][]{harchaoui2008kernel,li2015m}, but in our experience these are sensitive to the choice of kernels, and are thus not ideal for our context.}

\textcolor{black}{Even if the responses were not ordinal, the rationale for not applying CPD to both response and predictor variables is that the CPD algorithm does not account for the association between response and predictor variables, which is our main object of interest. For example, some initial change points may be for changes only in the response while there are no significant changes in the predictor variables. Since our algorithm aims to detect change points where the relationship between the response and predictors shifts and eventually, to help predict in the absence of the responses, the stress response variable was not included in the CPD algorithm.}
\end{remark}

\subsection{Step 2: Test the Significance of the \textcolor{black}{Co-Segments}}
\label{sse:step2}

Step 2 is crucial in this framework, as it aims to validate whether each segment of the response $Y$ and the predictor $X$ variables are associated. In other words, suppose that there is a change point in $X$. If $X$ and $Y$ are strongly associated, \textcolor{black}{a} change in $X$ would be highly likely to lead \textcolor{black}{a} change in $Y$. This indicates that change points in $X$ should \textcolor{black}{serve as potential candidate change points} in $Y$. Therefore, the objective of Step 2 is to verify the presence of an association between the passive sensing variables and the active variable at each \textcolor{black}{co-segment}.

At this step, typical self-reported active measures are discrete-valued \textcolor{black}{and ordinal} response variables. \textcolor{black}{To model the relationship in this case}, generalized linear models (GLMs) or other supervised learning methods can be used to \textcolor{black}{identify} the time points where different piecewise models \textcolor{black}{should be applied}. However, \textcolor{black}{not all co-segments provide a sufficient range of attainable values of responses}. For example, the stress level might \textcolor{black}{stay at 2 within a given segment but take 3 to 5 levels of values across the entire sample}. Since the model has to be fitted to each segment, this discrepancy can lead to model instability. \textcolor{black}{Consequently,} constructing GLMs for other categorical responses \textcolor{black}{is not preferred} \citep{clogg1995statistical, coxe2009analysis, warton2016three}. Such issues may arise even when wider ranging scales are \textcolor{black}{available}.

In our example, we \textcolor{black}{transform discrete-valued responses to be continuous by adding or subtracting additional scores to stress levels at each time $t$. In this study, we use Photographic Affect Meter (PAM) \citep{pollak2011pam} to construct the scores, which is described in Remark \ref{rem:adjustment}. By the transform,} we can consider linear associations instead of constructing non-linear equations. \textcolor{black}{For these linearized equations, we call continuous-valued response variable $\hat{Y}_t$ as the adjusted response, while $Y_t$ represents the original discrete response variable.}

For a fixed $k$th segment, we establish piecewise \textcolor{black}{linear equation $\hat{Y}_{t} = X_{t}'\beta_k + \epsilon_t$, $t\in\mathcal{T}_{k}$ and $k=1,\ldots,N_{\circ}$, where $\{\epsilon_t\}_{t\in\mathcal{T}_{k}}$ is the error with mean zero and constant variance in the $k$th segment.} Then, we test whether the equation from the segment is significant. The test of significance is equivalent to the test of nullity of the model, which can be performed by ANOVA. \textcolor{black}{For each $k$, the hypotheses become}
\begin{equation*}
    H_0: \quad \beta_{1,k} = \ldots = \beta_{p,k} = 0, \quad k=1,\ldots,N_{\circ},
\end{equation*}
and $H_1$ is not $H_0$. F-statistics can then be used to compute $p$-values for each $k$th segmentation. Based on the p-values of the F-tests, we merge the \textcolor{black}{co-segments of $\{X_t,Y_t\}$, $t\in\mathcal{T}_k$, whose association between the variables is not statistically significant with their neighboring shorter length segments. Note that in this step, the testing of nullity within each segment is irrelevant to the distinction between segments.} Denote $\tilde{\mathcal{T}}_k$ for $k=1,\ldots,\tilde{N}$ to represent the \textcolor{black}{subset of time indices after this refinement where $\tilde{N}$ is the number of the co-segments after Step 2, such that $\mathcal{L} = \uplus_{k=1}^{\tilde{N}}\tilde{\mathcal{L}}_k =\uplus_{k=1}^{\tilde{N}} \uplus_{t\in\tilde{\mathcal{T}}_{k}}\{X_t\}_{t}$}. We proceed with the algorithm if $\tilde{N}>1$.

\begin{figure}[t]
\centering
\includegraphics[width=0.45\columnwidth,height=0.4\textwidth]{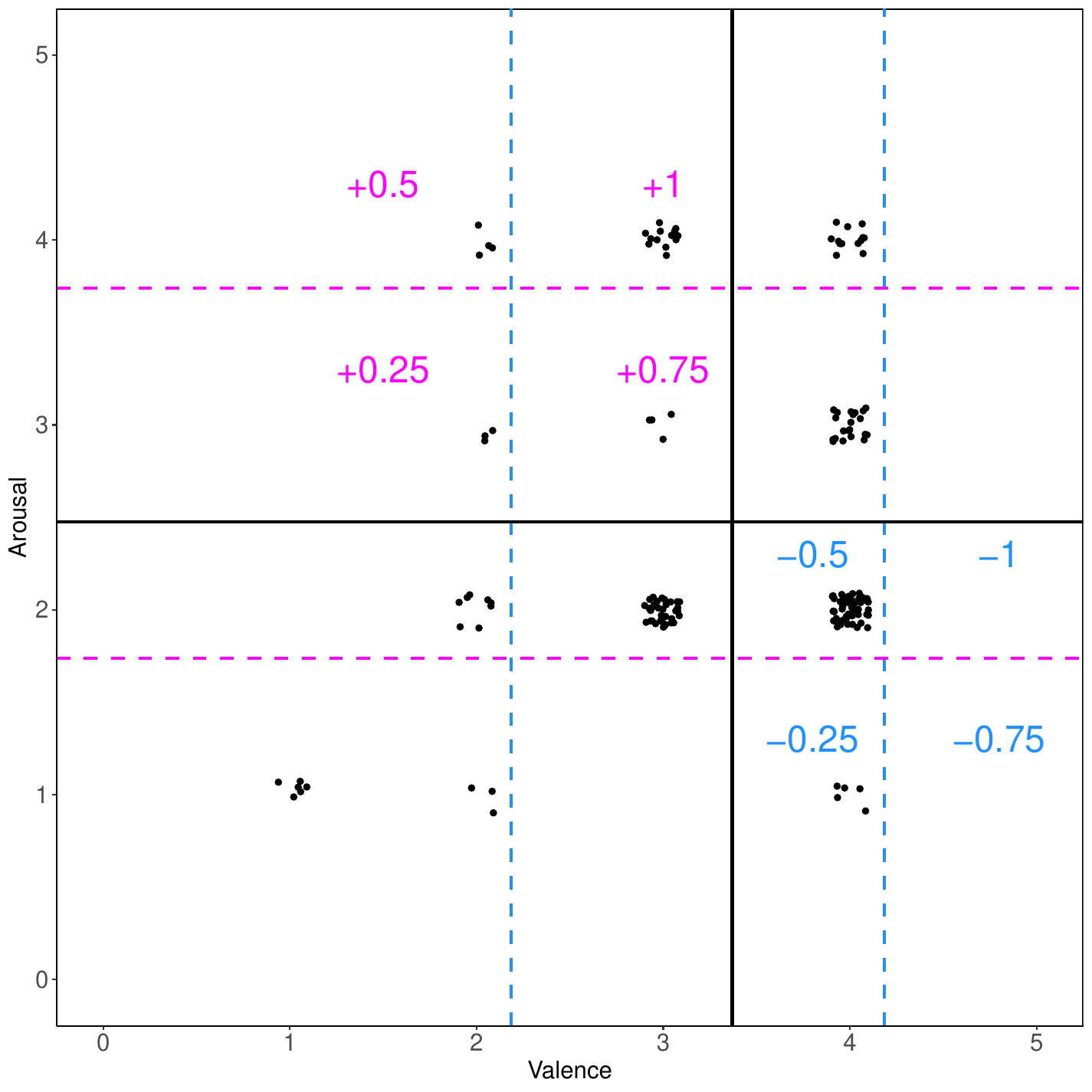}
\caption{Illustration of the derived PAM score \textcolor{black}{for the patient sternrelief5} from arousal and valence mapping onto Russell’s affective space. The annotated scores are added to the stress level at each time point to make the level continuous.}
\label{fig:pam_subj10}
\end{figure}

\begin{remark}\label{rem:adjustment}
\textcolor{black}{We illustrate how to construct adjusted responses $\hat{Y}_t$ from the 5-level stress indicator $Y_t$ with auxiliary information.} A unique feature of our application is the presence of two additional active variables; arousal and valence, that relate to our active variable of interest, i.e., emotional stress levels \citep{lang1990emotion,barrett1999structure}.  Arousal refers to the intensity of an emotion, ranging from calm or relaxed (value of 1) to excited or agitated (value of 5). In contrast, valence is the intrinsic pleasantness of an emotion, also ranging from \textcolor{black}{1  (unpleasant, sad) to 5 (pleasant, happy)}. The association between the measured intensities of arousal and valence with emotional stress is well known through EMA studies \citep{jefferies2008emotional,kuppens2013relation,bliss2020immutability}. In particular, we use the idea of the PAM \citep{pollak2011pam}, which maps the coordinates of arousal and valence onto Russell’s affective space \citep{russell1980circumplex}. Patients with high arousal (agitated) and low valence (unpleasant) are likely to have high emotional stress whereas patients with low arousal (relaxed) and high valence (pleasant) are likely to have low stress. We represent this in Figure \ref{fig:pam_subj10} which illustrates Russell’s affective space for a particular subject. The two solid black lines represent average arousal (range 1 to 5) and valence (range 1 to 5) and divide Russel's space into four quadrants with the top left quadrant indicating high stress and the bottom right quadrant indicating low stress. We further subdivide the top half (top left and top right quadrants) and the bottom half (bottom left and bottom right quadrants) with average arousal within those halves (dashed pink line). Similarly, we sub-divide the left half (top left and bottom left quadrants) and right half (top right and bottom right quadrants) with average valence within those halves (dashed blue line). The observed stress $Y_t$  in the top left quadrant of high stress is adjusted by adding a value of 0.25 to 1 (as indicated in the Figure \ref{fig:pam_subj10}) based on the sub-quadrant of their arousal and valence values and those in the bottom right quadrant of low stress are adjusted by subtracting a value of 0.25 to 1 similarly. \textcolor{black}{Note that this adjustment was specific to our illustrated example where values of arousal and valence were used to increase the variability in self-reported stress. Self-reported stress in our data was measured on a Likert scale and had low variability. This adjustment may not be necessary for other outcomes or other studies where the response has high variability and can be used as a continuous variable.}
\end{remark}

\subsection{Step 3: Test for Similarity of Association between \textcolor{black}{Co-Segments}}
\label{sse:step3}

This step is to verify whether the \textcolor{black}{co-segments $\{X_t,Y_t\}_{t\in\tilde{\mathcal{T}}_k}$ are distinct from their neighboring co-segments} $\{X_t,Y_t\}_{t\in\tilde{\mathcal{T}}_{k+1}}$. In other words, \textcolor{black}{if the relationships governing the two consecutive segments differ significantly, it suggests that a change in the association has occurred, implying} the change in the passive sensing variables may \textcolor{black}{lead to} the changes in the active variable. Conversely, if the two associations from the two consecutive \textcolor{black}{co-segments} are significant but are not distinguishable, one \textcolor{black}{may} think a meaningful transition in the active variable has not occurred.

Since we posit the linear associations, we use the Chow test (e.g., Exercise 2.12 in \cite{hayashi2011econometrics}) to $\tilde{N}-1$ pairs of the equations. Specifically, for $Y_t = X_{t}\beta_k + \epsilon_t$, $t\in\tilde{\mathcal{T}}_k$ and $Y_{t} =X_{t}\beta_{k+1} + \epsilon_t$, $t\in\tilde{\mathcal{T}}_{k+1}$, we test
\begin{equation*}
    H_0: \quad \beta_{k} = \beta_{k+1}, \quad k=1,\ldots,\tilde{N}-1,
\end{equation*}
and $H_1$ is not $H_0$ for $k=1,\ldots,\tilde{N}-1$. Similar to ANOVA in Step 2, F-statistics can be used to compute $p$-values for each pair of $k$th and $(k+1)$th segmentations. After performing all $\tilde{N}-1$ tests, merge the pairs of segments $\{X_t,Y_t\}$ that were determined to have similar associations. Denote $\hat{\mathcal{T}}_k$ for $k=1,\ldots,\hat{N}$ to represent the \textcolor{black}{subset of time indices after this refinement, such that $\mathcal{L} = \uplus_{k=1}^{\hat{N}}\hat{\mathcal{L}}_k =\uplus_{k=1}^{\hat{N}} \uplus_{t\in\hat{\mathcal{T}}_{k}}\{X_t\}_{t}$, where $\hat{N}$ is the number of the co-segments after Step 3}. We proceed with the algorithm if $\hat{N}>1$.

\subsection{Step 4: Test for the Similarity of the Distributions of the Responses}
\label{sse:step4}

We \textcolor{black}{assess} whether the distributions of $\{Y_t\}$ are similar to its neighboring segments. This step \textcolor{black}{builds on} the co-segment result from the previous steps. Specifically, in Steps 1-–3, we validate changes in patterns of the passive sensing variables, the significance of changes in associations between passive and active variables, and the similarities between co-segments. We empirically find that, even when the discoveries in the previous steps are statistically significant, there are \textcolor{black}{instances} where meaningful changes in the patterns of active variables \textcolor{black}{are not} observed. This suggests that changes in physical activity, sociability, and sleep patterns measured by the sensors may not \textcolor{black}{always} induce meaningful changes in emotional stress states. Therefore, in this step, we validate whether the patients feel the changes in their status aligned with what the algorithm has discovered.

To determine whether there is a significant change between two consecutive segments of the response variable $\{Y_{t}\}_{t\in\hat{\mathcal{T}}_k}$ and $\{Y_{t}\}_{t\in\hat{\mathcal{T}}_{k+1}}$, $k=1,\ldots,\hat{N}-1$, we can conduct tests of equivalence of proportions between the two samples. If only two levels are present, we can perform a $Z$-test. Otherwise, if there are more than three observed levels, we can use the Chi-squared test. After performing all $\hat{N}-1$ tests, merge the \textcolor{black}{co-segments of $\{X_t,Y_t\}$} that were determined to be indistinguishable. Let $\mathcal{T}_k^*$ represent the set of time indices after this refinement \textcolor{black}{and $N^*$ be the number of the co-segments after Step 4}. We proceed with the algorithm if $N^*>1$.

\begin{remark}\label{rem:complicated}
Through Steps 2 - 4, we have ruled out the initial \textcolor{black}{co-segments where the} association or distribution of the responses is similar to their neighbors. \textcolor{black}{While} we focus on introducing the algorithm so that the simplest approach at each step has been proposed, each step can be \textcolor{black}{adapted or replaced with alternative methods}. For instance, if GLM is considered in Step 2 instead as explained in Section \ref{sse:step2}, a likelihood ratio test between the saturated model and the null model can be performed instead of ANOVA. The algorithm for testing structural breaks in GLMs \citep{zeileis2006implementing} can also replace the Chow test in Step 3. Finally, \textcolor{black}{in Step 4, rather than} comparing \textcolor{black}{sample} proportions, one can employ goodness-of-fit tests of cumulative distribution functions using criteria such as the Kolmogorov-Smirnov or Cramér–von Mises criterion (e.g., Chapter 19.3 in \cite{van2000asymptotic}) at Step 4. 
\end{remark}

\subsection{Step 5: Label Prediction as Classification Problem}
\label{sse:step5}

Once the algorithm has proceeded with Steps 1-4, we assign ordinal class labels to the finalized segments of $\{X_t\}$. For example, class label 1 is assigned to the first segment, class label 2 to the second segment, and so on. This labeling is justified since we have ensured that the consecutive co-segments of $\{X_t,Y_t\}$ are distinct from their neighboring segments. \textcolor{black}{Note that the labels assigned are ordinal and do not have clinical meanings; one} can simply think that the assigned class labels are assumed to represent different emotional stress states. \textcolor{black}{For example, if two labels are assigned, the patient is regarded as in a stress period during the first segment, while he or she is not in the stress period during the second segment, or vice versa. Mathematically, let $\{S_t\}_{t\in\mathcal{T}_k^*}$, $k=1,\ldots,N^*$, be a set of class labels, where $S_t \in \{1,2,\ldots,N^*\}$. For instance, if $N^*=2$, we can assign $S_t = 1$, $t\in\mathcal{T}_1^*$, is the label for the period when the patient is under the emotional stress period if the mean of $\{Y_t\}_{t\in\mathcal{T}_1^*}$ is larger than that of $\{Y_t\}_{t\in\mathcal{T}_2^*}$. Further investigation should be involved to connect the labels with the clinical status of emotional stress status.} With the assigned labels, we can solve a classification problem for obtaining $\hat{S}_t$, $t>T$, from the pairs of $\{S_t,X_t\}_{t\in\mathcal{T}}$, $\mathcal{T} = \uplus_{k=1}^{N^*} \mathcal{T}_{k}^*$.

For classification methods, we employ logistic (multinomial) regression (Reg), linear discriminant analysis (LDA), standard support vector machine (SVM), random forests (RF), and gradient boosting (Boost) for $\{S_t,X_t\}_{t\in\mathcal{T}}$. The last three methods are available in well-known R packages \texttt{e1071} \citep{dimitriadou2009package}, \texttt{randomForest} \citep{liaw2002classification}, and \texttt{gbm} \citep{ridgeway2004gbm}, and the practical implementation of all five methods is described in James et al. \citep{james2013introduction}. Note that more complicated classification methods could be also considered at this step.

\begin{remark}\label{rem:dichotomized}
\textcolor{black}{While we did not observe such a case in our data, it is certainly possible to merge finalized segments when too many segments are produced. A basic approach is to modify Step 4 to examine whether co-segments that are not direct neighbors have distinct mean stress levels. Alternatively, Step 4 can be adjusted to assess more comprehensive hypothesis tests for dissimilarity in responses. For example, a test of equal variances could be added to Step 4. Test of equal distributions by the methods discussed in Remark \ref{rem:complicated} can be also considered.}
\end{remark}

%%%%%%%%%%%%%%%%%%%%%%%%%%%%%%%%%%%%%%%%%%%%%%%%%%%%%%%%%%%%%%%%%%%
\section{Application}
\label{se:application}

\subsection{Illustration of the Algorithm}
\label{sse:illustration}

In this section, we provide an illustrative example focusing on one specific patient, identified as \textcolor{black}{sternrelief5} within our dataset. The result of the algorithm application for another patient \textcolor{black}{sternrelief9, for whom the number of finalized segments is different,} will be presented in Appendix \ref{sap:illustration}.

\begin{figure}[t]
\centering
\includegraphics[width=1\columnwidth,height=0.7\textwidth]{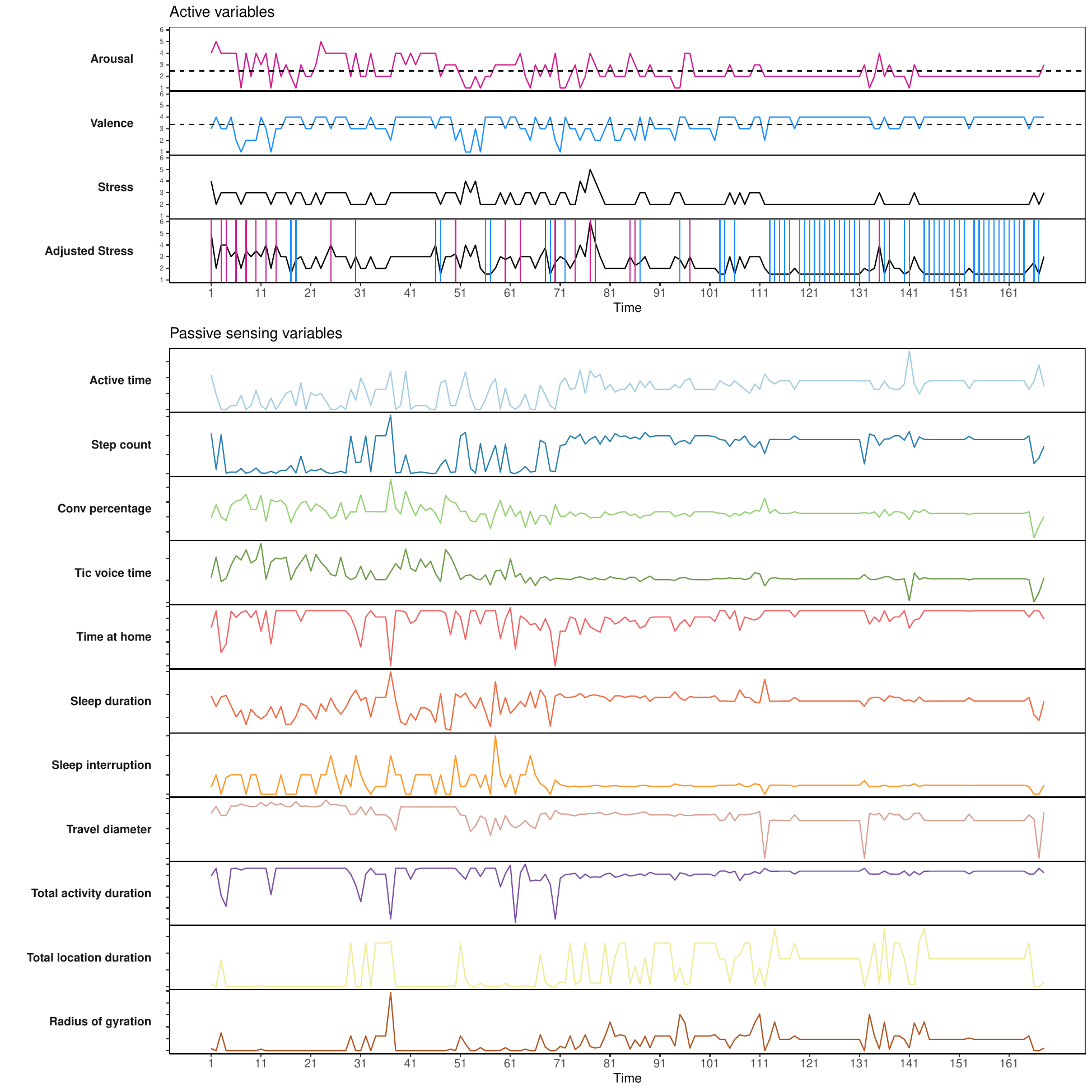}
\caption{Time plots of the stress level, arousal, valence, and adjusted stress level (top panel) and the 11 passive sensing variables for the patient \textcolor{black}{sternrelief5} (bottom panel). The description of the variable used in the plots can be found in Table \ref{tab:data}.}
\label{fig:initialization1}
\end{figure}

Figure \ref{fig:initialization1} displays the initial data for patient \textcolor{black}{sternrelief5}, along with the adjusted stress level as described in Remark \ref{rem:adjustment} in Section \ref{sse:step2}. In the top panel, the first four series represent the response variables, including arousal, valence, stress level, and adjusted stress level. Note that arousal and valence often change simultaneously but in opposite directions, indicating a negative correlation between the two items. Therefore, our analysis focuses on the second and fourth quadrants of Russell’s affective space described in Remark \ref{rem:adjustment}. For the adjusted stress level, the red and blue lines on the adjusted stress level denote the timing when the patient experiences \textcolor{black}{high values of arousal and low values of valence, and low values of arousal and high values of valence}, respectively. \textcolor{black}{The red lines are relatively dense at the beginning, while the blue lines are sparse. Over time, the blue lines become denser, whereas the red lines become sparse toward the end of the observation period. Interestingly, this pattern corresponds to a gradual downward shift in the mean levels.} In the bottom panel, the 11 series illustrate the passive sensing variables used as predictors. Similar patterns of changes can be observed across various passive sensing variables around the time points where the densities of the red and blue lines are changing.

\begin{figure}[t]
\centering
\includegraphics[width=1\columnwidth,height=1\textwidth]{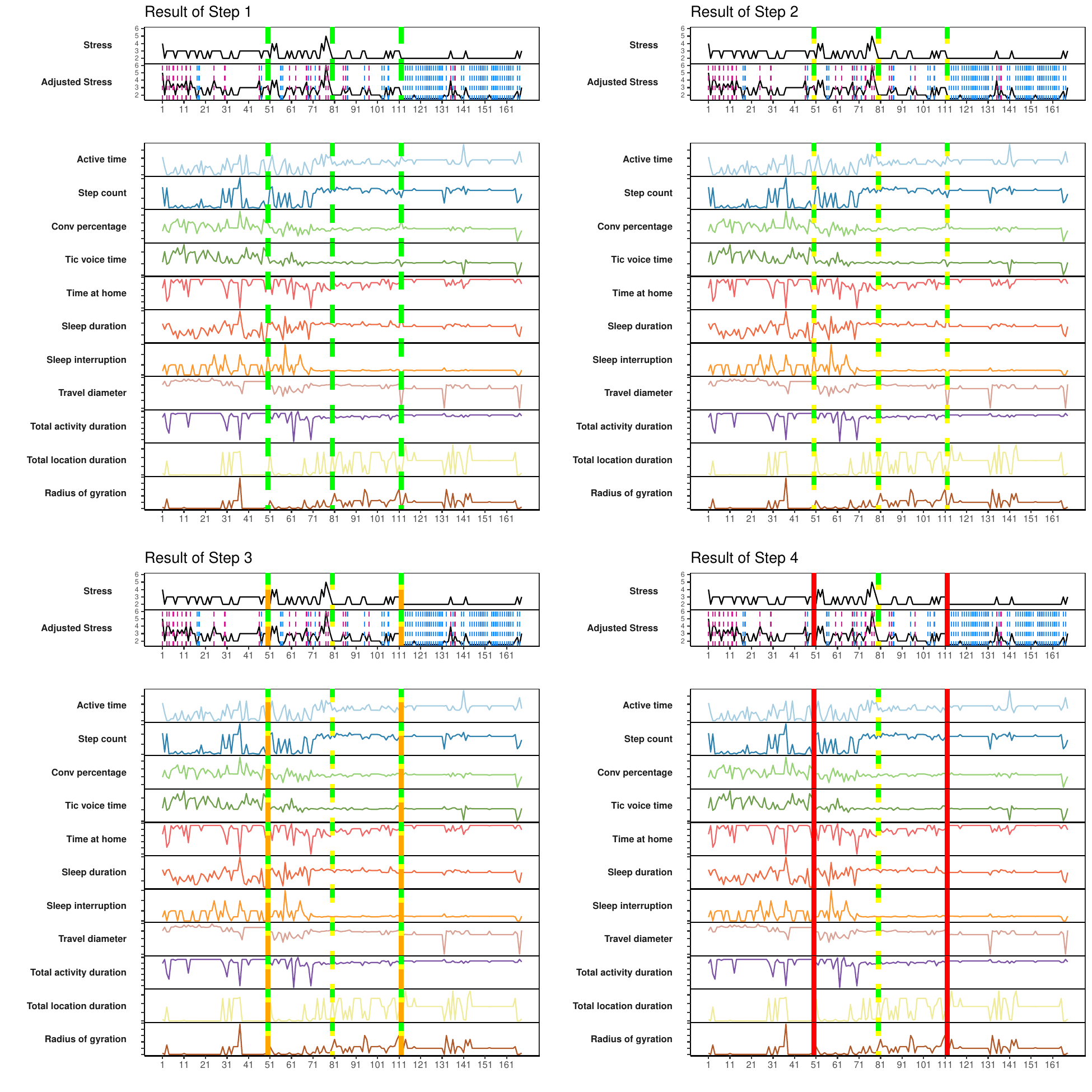}
\caption{Illustration of the results for the patient \textcolor{black}{sternrelief5} after each step. The top left panel represents the result after Step 1. The horizontal green dot-dashed line indicates the initial change points at $t^*=50,80,112$. The top right panel shows the result after Step 2, where the validated time point overlaps the result in Step 1 and is marked as a thick yellow dotted line. The bottom two panels indicate the result after Steps 3 and 4, which are marked as orange dashed and red solid lines, respectively.}
\label{fig:result_illustration1}
\end{figure}

Figure \ref{fig:result_illustration1} depicts the sequential outcome of each step of the algorithm for patient sternrelief9, as described in Sections \ref{sse:step1} -- \ref{sse:step4}. In the top left panel, the result of Step 1 shows that \textcolor{black}{three change points $t^*=50,80,112$ were} estimated, highlighted by \textcolor{black}{thick green dot-dashed lines. This forms $N_{\circ}=4$ co-segments, indicating} that the CPD algorithm detects significant changes in the mean levels or variances of the passive sensing variables before and after \textcolor{black}{those 3 initially} estimated times. \textcolor{black}{Interestingly, the densities of the red and blue lines vary significantly across segments although the information of the active variables are not involved in this step.} The top right panel displays the result after Step 2, where the estimated change point is confirmed by testing the significance of the linear relationship between the adjusted stress level and the passive sensing variables in each segment. \textcolor{black}{From $N_{\circ} = 4$, four tests of nullity were performed, and all linear associations between the passive sensing variables and the adjusted stress levels from the co-segments were found to be significant $(\tilde{N}=4)$. The consequence of Step 2 are drawn as thick yellow dotted lines, which overlap with the green lines from the result of Step 1. Note that the plot of the adjusted stress level reveals more blue lines in the first segment compared to the second. This corresponds to the tendency of the stress levels to be high in the second segment while it is relatively low in the first segment.}

\textcolor{black}{In the bottom left panel of Figure \ref{fig:result_illustration1}, the orange dashed lines represent the results of Step 3, confirming the separation of consecutive co-segments. Three sequential hypothesis tests were performed, and the results indicated no statistically significant distinction between the second and third co-segments. As a result, these co-segments were merged, leaving three co-segments ($\hat{N}=3$) after Step 3.} Finally, the result of Step 4 validates whether the distribution of stress levels in the \textcolor{black}{pairs of segments are different. The finalized number of co-segments is $N^*=3$, and the each of segment of $\{X_t\}_{t\in\mathcal{T}_k^*}$, $k=1,2,3$, will have different class labels, $S_t=1,2,3$. Interestingly, the rejections of the null hypotheses seem related to the differences in the densities of the blue and red solid lines in each segment of the adjusted stress levels}, which implies that the patient also finds a subjective change in the level of stress. Time plots and illustrations of results for another patient, \textcolor{black}{sternrelief9, are provided in Appendix \ref{sap:illustration}. This patient has been chosen to illustrate another case where the initial estimated change point $(N_{\circ}=1)$ at $t^*=69$ turns out to be the finalized estimated change point $(\tilde{N}=\hat{N}=N^*=1)$ through the rest of the steps,} which can be confirmed by Figure \ref{fig:result_illustration2}.

\clearpage

\clearpage
\subsection{Experiment Design}
\label{sse:design}

Since the data has been collected in a natural environment, we do not know the emotional stress status of the patients. \textcolor{black}{Therefore}, we consider the algorithm's results based on each patient's full sample, as illustrated in Section \ref{sse:illustration}, to serve as the ground truth. Specifically, \textcolor{black}{if $N^*=2$}, with one having a higher mean stress level, we assume the patient was in a stress period during that segment and not during the other. \textcolor{black}{To evaluate performance, we withhold subsamples as test data,} allowing the algorithm to predict labels for these samples based on segmentation results from the remaining dataset. This approach serves as a measure of the algorithm's robustness and is akin to assessing in-sample performance.

The data generation process is structured as follows. For each patients, we start selecting 5 consecutive samples from $t=6$, so that the subsamples are from $t=6,\ldots,10$ out of the total time points $\mathcal{T} =\{1,\ldots,T\}$ in the initial iteration. These held-out samples constitute the test data, denoted by $\mathcal{T}_{\textrm{test}}$, while the remaining samples consist of the training data, denoted by $\mathcal{T}_{\textrm{train}}$. We then apply Steps 1-4 of the stress period detection algorithm to the training data. \textcolor{black}{We consider the procedure a success} if the procedure goes through all steps without a halt. If the algorithm \textcolor{black}{suceeds}, we count this iteration as a success and proceed to execute Step 5. However, if the algorithm fails to complete any of the steps, we move on to the next set of 5 consecutive samples at $t=6+s,\ldots,10+s$, where $s$ denotes the step size of the moving window, \textcolor{black}{using} a rolling-window strategy. We consider $s=2$ and $5$. If the moving window reaches one of the finalized change points described in Section \ref{sse:illustration}, we skip this iteration and proceed to the next window. This process continues until we achieve 50 successes or until the largest $t$ in $\mathcal{T}_{\textrm{test}}$ reaches the final observation time point $T$. By following this strategy, and using the known $N^*$ values and segment results (2 for 3 selected patients, 3 for the other 3 selected patients), the test data held out from the first segment has a class label of 1. Similarly, class labels 2 and 3 can be assigned. For example, from the result of patient \textcolor{black}{sternrelief5} in Figure \ref{fig:result_illustration1}, the \textcolor{black}{finalized estimated change points were $t^*=50,112$. So, if the samples are chosen to be held out before $t=50$, the true label for this trial is 1. If the held out samples are between $t=50$ and $t=112$, the true label is regarded as 2. Otherwise, the true label is assigned to 3.} For the case of \textcolor{black}{sternrelief9, for instance, the possible labels assigned to the held out samples are 1 or 2, depending on the time indices of those samples are less than or greater than $t^*=69$.} This strategy enables the construction of true labels from the algorithm's results for all 6 patients, where the number of finalized segments and their corresponding time points are determined.

We explain several benchmarks against the proposed algorithm. The first approach involves using only the self-reported data. Specifically, let $\hat{Y}_{\textrm{test},t}$, $t \in \mathcal{T}_{\textrm{test}}$, and $\hat{Y}_{\textrm{train},t}$, $t \in \mathcal{T}_{\textrm{train}}$, denote the continuous response variables for the test data and the training data, respectively. This segment is extracted solely from the adjusted stress level computed using all available samples. Then, utilizing the results from the full data, we apply K-means clustering to $\hat{Y}_{\textrm{train},t}$ with two (or three) centers of the clusters. Denote the means of the two clusters as $\bar{Y}_k$, $k=1,2$ if $N^*=2$ (or $k=1,2,3$ if $N^*=3$). The label can then be determined by minimizing the squared difference,
\begin{equation}\label{e:only_y}
    s_{\textrm{test},t} = \argmin_{k=1,\ldots,N^*}(\hat{Y}_{\textrm{test},t} - \bar{Y}_k)^2, \quad t\in\mathcal{T}_{\mbox{test}}.
\end{equation}
We \textcolor{black}{refer to} this labeling \textcolor{black}{as} response only (Resp). Another benchmark we consider \textcolor{black}{involves} constructing the relationship between the response variable and the predictor variables without segmentation. In this approach, we use the model $\hat{Y}_{t} = X_t' \hat{\beta}_{\textrm{train}} + \epsilon_t$, $t \in \mathcal{T}_{\textrm{train}}$, \textcolor{black}{with the same assumption of $\{\epsilon_t\}$ in Section \ref{sse:step2}}. Here, $\hat{Y}_t$ represents the adjusted stress level, and $X_t$ denotes the passive sensing variables. \textcolor{black}{Then the predicted responses with $\hat{\beta}_{\textrm{train}}$ can be computed as}
\begin{equation*}
    \hat{\hat{Y}}_{\textrm{test},t} = X_{\textrm{test},t}'\hat{\beta}_{\textrm{train}}, \quad t\in\mathcal{T}_{\textrm{test}}.
\end{equation*}
Thus, the label in this approach can be determined by finding the minimizer of the squared difference,
\begin{equation}\label{e:regression}
    s_{\textrm{test},t} = \argmin_{k=1,\ldots,N^*}(\hat{\hat{Y}}_{\textrm{test},t} - \bar{Y}_k)^2, \quad t\in\mathcal{T}_{\mbox{test}},
\end{equation}
where $\bar{Y}_k$ are the means of the clusters used in \eqref{e:only_y}. We call this labeling \textcolor{black}{as} prediction (Pred). Two benchmarks \eqref{e:only_y} and \eqref{e:regression} is to \textcolor{black}{determine} whether the passive sensing variable is informative to infer the state of the emotional stress. \textcolor{black}{In addition}, the difference between the linear classifiers, Reg and SVM, and Pred is that the former considers the segmentation. While the proposed methods construct piecewise linear relationships in each segment, whose significance and separation are proven by the algorithm, Pred does not leverage the segment results. \textcolor{black}{Furthermore}, note that the squared difference in \eqref{e:only_y} can be adjusted to the average over $t$, which means $s_{\textrm{test}} = \argmin_{k=1,\ldots,N^*}\sum_{t\in\mathcal{T}_{\mbox{test}}}\frac{1}{5}(Y_{\textrm{test},t} - \bar{Y}_k)^2$, where a similar \textcolor{black}{variation} can be applied to that in \eqref{e:regression}. However, \textcolor{black}{this variation complicates the comparison of the result} with the results produced by the classifiers. Therefore, we adhere to the pointwise scheme.

% https://www.evidentlyai.com/classification-metrics/multi-class-metrics
For evaluation, we define performance measures as follows. We treat the 5 consecutive samples in the test data as independent trials, resulting in 5 times the number of successful iterations in this experiment, \textcolor{black}{which provides} sufficient trials for analyses. First, we consider macro-averaging precision and recall in the context of the multi-class classification problem. Specifically, for $k$th class label, \textcolor{black}{recall of the $k$th class label can be computed as}
\begin{displaymath}
    \mbox{Rec}_k = \frac{\textrm{Correct class $k$ prediction}}{\textrm{All class $k$ instances}}, \quad k=1,\ldots,N^*.
\end{displaymath}
Then the recall is computed as
\begin{displaymath}
    \mbox{Rec} = \frac{\mbox{Rec}_1 + \ldots + \mbox{Rec}_{N^*}}{N^*}.
\end{displaymath}
Similarly, for precision of the $k$th class
\begin{displaymath}
    \mbox{Prec}_k = \frac{\textrm{Correct class $k$ prediction}}{\textrm{All class $k$ prediction}}, \quad k=1,\ldots,N^*,
\end{displaymath}
the precision, denoted by $\mbox{Prec}$, can be computed similarly
\begin{displaymath}
    \mbox{Prec} = \frac{\mbox{Prec}_1 + \ldots + \mbox{Prec}_{N^*}}{N^*}.
\end{displaymath}
Note that the performance measures mentioned above remain applicable even when the number of final segments is only 2. For instance, if the labels `0' and `1' in both the actual and predicted classifications were interchanged, the precision, false discovery rate, F1 score, and Matthews correlation coefficient would change accordingly. Since determining which class label should be designated as `0' can be subjective, we opt to employ the same performance measures typically used for multi-class classification problems in our illustrative example. We also report accuracy, denoted by $\mbox{Acc}$,
\begin{displaymath}
    \mbox{Acc}
    = \frac{\textrm{Correct class $1$ prediction}+\ldots+\textrm{Correct class $N^*$ prediction}}{\textrm{All class $1$ prediction}+\ldots+\textrm{All class $N^*$ prediction}}.
\end{displaymath}
Note that it is known that the micro-averaging recall, precision, and accuracy are all equal. Finally, we also consider F1-score $(\mbox{F1}_k)$,
\begin{displaymath}
    \mbox{F1}_k 
    = \frac{2 \times \mbox{Rec}_k \times \mbox{Prec}_k}{\mbox{Rec}_k + \mbox{Prec}_k}, \quad k =1,\ldots,N^*.
\end{displaymath}
Other performance measures, for instance, false positive rate, type-I or type-II errors, etc can be also considered. However, false discovery rate and type-II error can be derived from the precision and recall, respectively. More importantly, depending on which class is selected to be 0, the quantities of type-I and type-II would be swapped, so that in the multi-class classification problem, those two quantities would indicate the same result. For this reason, \textcolor{black}{and in the interest of} concise representation in the tables below, we only report the F1-score, a harmonic average of the recall and precision, for each $k$th class.

\subsection{Experiment Results}
\label{sse:results}

The summary of the performance measures, \textcolor{black}{for each} data generation and true label construction described in Section \ref{sse:design}, is shown in Table \ref{tab:result} \textcolor{black}{in Appendix \ref{sap:simulation}.}

Overall, among the methods considered, the RF and Boost consistently outperform the \textcolor{black}{others} across all patients in most performance measures. Specifically, compared to the benchmarks, the \textcolor{black}{superior performance of segments suggests that incorporating} piecewise segmentation \textcolor{black}{improves the understanding of} the association between response and passive sensing variables. \textcolor{black}{Furthermore, the stronger performance of RF and Boost} compared to other classifiers \textcolor{black}{indicates that employing more sophisticated models, rather than simple linear relationships, enhances the identification of} these associations. Between the benchmarks, Pred does not outperform Resp overall, \textcolor{black}{suggesting} that incorporating passive sensing variables is advantageous only when the segmentation \textcolor{black}{is taken into account}. This \textcolor{black}{finding} implies that the \textcolor{black}{relationship} between the stress levels and passive sensing variables \textcolor{black}{evolves across} segments, \textcolor{black}{reinforcing the motivation} for introducing segmentation described in Section \ref{se:intro}.

Among recall, precision, and accuracy, most methods \textcolor{black}{yield} similar scores, \textcolor{black}{indicating} that, on average, the algorithm predicts class labels with decent quality. However, when comparing the F1 scores for each class label, substantial discrepancies can be observed across all cases. In particular, Reg and SVM \textcolor{black}{fail} in some cases, \textcolor{black}{suggesting} that \textcolor{black}{these methods} tend to produce biased results, which need to be addressed.

Finally, compared to the results from the patients with $N^{*}=2$, the results from the last three cases \textcolor{black}{appear} worse. \textcolor{black}{This decline} is fully \textcolor{black}{expected, as} multi-label classification problems \textcolor{black}{are obviously} more challenging. \textcolor{black}{However,} the magnitude of the \textcolor{black}{performance} drop is less severe, in terms of accuracy \textcolor{black}{as} most proposed methods \textcolor{black}{maintain scores} above 1/3, \textcolor{black}{and similar trends are observed for} recall and precision.

%%%%%%%%%%%%%%%%%%%%%%%%%%%%%%%%%%%%%%%%%%%%%%%%%%%%%%%%%%%%%%%%%%%
\section{Conclusion}
\label{se:conclusion}

In this study, we proposed a data-driven approach for detecting the emotional stress periods of patients. Two types of mHealth data are used: personal passively measured data and self-reported surveys. The active variables \textcolor{black}{include} stress levels (the response variable) and related \textcolor{black}{emotions, such as} arousal and valence. While conventional algorithms rely on physiological sensing data, our proposed algorithm \textcolor{black}{is innovative in that it works with} different types and combinations of mHealth data that have not been previously analyzed. In addition to the unique \textcolor{black}{data combinations, we apply} the concept of segmentation to \textcolor{black}{explore where the simultaneous changes in both data types are associated.}

Throughout the study, we illustrated and explained the initialization and the results \textcolor{black}{at} each step of the algorithm. \textcolor{black}{The primary goal of} the algorithm \textcolor{black}{was} to estimate change points in the passively sensed data and then confirm the changes in the stress level, \textcolor{black}{ensuring these changes aligned with the shifts in patterns}. The algorithm finalized the estimated change points by \textcolor{black}{examining} the separation of the \textcolor{black}{co-segments} by checking distinct associations and \textcolor{black}{difference in the} mean levels of the response variable. Class labels were assigned to the segments, and the segmented passive sensing variables were used to predict the class label for newly inflowing data. We considered multiple patients and \textcolor{black}{focused on} one case. From the series of plots, we visualized relationships between stress levels and related measures, and passive sensing variables. \textcolor{black}{To validate the algorithm,} we proposed an experiment \textcolor{black}{that included} data generation, performance \textcolor{black}{metrics}, and benchmarks. The results of \textcolor{black}{the validation demonstrated} that the proposed approach outperformed the benchmarks, \textcolor{black}{which aligned} with our expectations.

\textcolor{black}{There are limitations to our approach. First, we include patients who have 50 days or more of passive and active data. The length of the period of observation would depend on the signal-to-noise ratio of the data, but a sufficient period of observation is needed to get meaningful insights into our experiment. Second, our algorithm is set for the case where passive and active measures are measured at the same frequency i.e., daily. Typically, passive measures can be recorded more frequently (e.g., intraday) and we suggest aggregating passive measures to the frequency of the response or active measure. Third, our algorithm requires arousal and valence to augment self-reported stress to increase variability in the self-reported stress measure, which may not be readily available in other studies measuring stress. Fourth, we have not controlled for multiple comparisons, and acknowledge that controlling for global type 1 error in this algorithm is complex and needs further research. Lastly, changes in other moments, for example, autocorrelations of the passively sensed measures, are not considered in our work.}

There are several possible extensions of the current research. First, one \textcolor{black}{could} replace the piecewise linear relationships between the stress level and the passive sensing variables with other non-linear models. Another extension \textcolor{black}{would incorporate} demographic and behavioral variables in the \textcolor{black}{approach}. The motivation is that it is \textcolor{black}{generally challenging} to collect a sufficient \textcolor{black}{number} of samples for each individual. If \textcolor{black}{similarities in relationships within the same grouped individuals can be modeled, it may be beneficial to leverage} samples \textcolor{black}{from} other individuals in the group for the target patient. Lastly, it \textcolor{black}{would also be} interesting to develop a unique change point detection algorithm that can be employed in the dataset we use. In particular, our task is \textcolor{black}{akin to identifying} change points in a series of non-linear equations. To our knowledge, however, \textcolor{black}{no such algorithm has been developed} yet.

%\backmatter
\section*{Acknowledgements}

S Banerjee acknowledge partial supports from NIMH award P50MH113838 and NIA award P01AG073090. YK and S Basu acknowledge partial support from NSF CAREER award DMS-2239102. In addition, S Basu acknowledges partial support from NSF awards DMS-1812128, DMS-2210675, and NIH awards R01GM135926, R21NS120227.

\section*{Data Availability Statement}
The R code used in Section \ref{se:application} is available on GitHub at \href{https://github.com/yk748/SPD}{https://github.com/yk748/SPD}. The data that support the findings of this study are available on request from the corresponding author. The data are not publicly available due to privacy or ethical restrictions.

\section*{Conflict of Interest}
The authors have declared no conflict of interest.

\small
\bibliography{spd_arxiv}

\newpage
\appendix

\section*{Appendix}

\section{Overview of the Algorithm}
\label{sap:overview}
\vspace*{12pt}

\begin{figure}[h]
\centering
\includegraphics[width=0.6\textwidth, height=0.7\textheight]{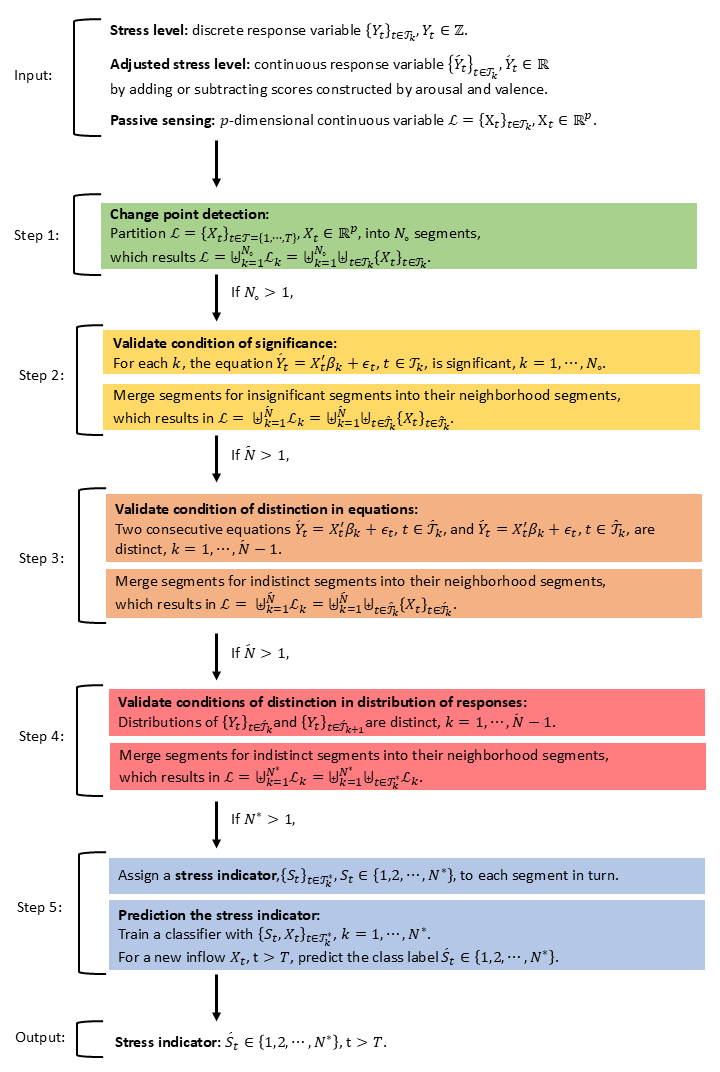}
\caption{Flowchart of predicting emotional stress states algorithm. The details for each step can be found in Sections \ref{sse:step1} -- \ref{sse:step5}.}
\label{fig:flowchart}
\end{figure}

\newpage
\section{Description of data in the analyses}
\label{sap:data}
\vspace*{12pt}

% Please add the following required packages to your document preamble:
% \usepackage{graphicx}
\begin{table}[h]
\resizebox{\textwidth}{!}{%
\begin{tabular}{llll}
\hline
\multicolumn{1}{c}{Name}           & \multicolumn{1}{c}{Type} & \multicolumn{1}{c}{Source} & \multicolumn{1}{c}{Description} \\ \hline
Active time                        &  Continuous    &  Pedometer      &  Total time in seconds a person is walking during the day.           \\ \cline{1-1}
Step count                          &  Continuous    &  Pedometer      &  Step counts in a day.           \\ \cline{1-1}
Conversation percentage           &  Continuous    &  VoicePresence      &  Percentage of time that voice sensing is running in which voice is detected.           \\ \cline{1-1}
Tic voice time                      &  Continuous    &  VoicePresence      &  Time length voice detected during activities.           \\ \cline{1-1}
Time at home                        &  Continuous    &  AppDailyReport      &  Percentage of a day a person spends at home in a 24 hour period            \\ \cline{1-1}
Sleep duration                   &  Continuous    &  AppDailyReport      &  Number of seconds a person slept.           \\ \cline{1-1}
Sleep interruption                &  Continuous    &  AppDailyReport      &  Number of periods of non-stationary data from the Activity stream during the inferred sleep period.           \\ \cline{1-1}
Travel diameter                     &  Continuous    &  AppDailyReport      &  Total diameter in meters a person travels in a day.            \\ \cline{1-1}
Total activity duration             &  Continuous    &  Activity      &  Total uptime of the Activity stream.           \\ \cline{1-1}
Total location duration            &  Continuous    &  Activity      &  Total uptime of the Location stream.           \\ \cline{1-1}
Radius of gyration               &  Continuous    & Daily Summary       &  Measure that relates to the area covered by a user.           \\ \hline
Stress                            &  Ordinal    &  Survey      &  level of feeling emotional stress           \\ \cline{1-1}
Arousal                           &  Ordinal    &  Survey      &  Level of autonomic activation that an event creates, and ranges from calm (or low) to excited (or high).            \\ \cline{1-1}
Valence                          &  Ordinal    &  Survey      &  Level of pleasantness that an event generates and is defined along a continuum from \textcolor{black}{low to high}.    \\ \hline
\end{tabular}%
}
\caption{Data dictionary of Alacrity Phase I used in the study. Type stands for the type of variables, Source represents how the corresponding data has been collected, and Description contains the short description of each variable.}
\label{tab:data}
\end{table}

\section{Description of Matteson and James (2014)}
\label{sap:matteson}

\textcolor{black}{Here we outline the change point detection (CPD) algorithms proposed by \cite{matteson2014nonparametric}. The algorithm consists of two parts; estimation of change point location and hierarchical significance testing.}

\textcolor{black}{The estimating change point relies on the approximation of divergence measure \citep{szekely2005hierarchical}, a distance between multivariate distributions computed by their characteristic functions. Suppose $X'$ and $Y'$ are independent copies of $X$ and $Y$, respectively. The divergence measure based on Euclidean distances with $\alpha\in(0,2)$ is defined as
\begin{displaymath}
    \mathcal{E}(X,Y;\alpha) = 2\mathbb{E}|X-Y|^{\alpha} - \mathbb{E}|X-X'|^{\alpha} - \mathbb{E}|Y-Y'|^{\alpha}.
\end{displaymath}
Its empirical analog can be obtained based on U-statistics; for $n$ and $m$ independent copies $\mathbf{X}_n=\{X_{i}\}_{i=1,\ldots,n}$ and $\mathbf{Y}_m=\{Y_{j}\}_{i=1,\ldots,m}$ of $X$ and $Y$, respectively, the empirical divergence measure is defined as
\begin{equation}\label{e:u-statistics}
    \hat{\mathcal{E}}(\mathbf{X}_{m},\mathbf{Y}_{m}(\kappa);\alpha) 
 = \frac{2}{m+n}\sum_{i=1}^n\sum_{j=1}^m|X_i-Y_j|^{\alpha} - \binom{n}{2}^{-1}\sum_{1\leq i < k\leq n}|X_i - X_k|^{\alpha} - \binom{m}{2}^{-1}\sum_{1\leq j < k\leq m}|Y_j - Y_k|^{\alpha}.
\end{equation}
The algorithm uses the U-statistic \eqref{e:u-statistics} to identify the location. Suppose that a sequence of observations $\{Z_t\}_{t=1,\ldots,T}$ is given. Assume that $k-1$, $k\geq1$, change points $\hat{\tau}_0 = 0<\hat{\tau}_1<\ldots<\hat{\tau}_{k-1}<\hat{\tau}_k=T$ have been identified. For $k$ segmented observations $\hat{\mathcal{C}}_{i}=\{Z_{\hat{\tau}_{i-1}+1},\ldots,Z_{\hat{\tau}_i}\}$, $i=1,\ldots,k$, the new change point $\hat{\tau}(i)$ within $i$th cluster, $i=1,\ldots,k$, is identified by 
\begin{equation}\label{e:changepoint}
    (\hat{\tau}(i),\hat{\kappa}(i)) 
    = \mathop{\rm argmax}_{(\tau,\kappa)} \hat{\mathcal{Q}}(\mathbf{X}_{\tau(i)},\mathbf{Y}_{\tau(i)}(\kappa(i));\alpha),
\end{equation}
where for $\mathbf{X}_{\tau(i)} = \{Z_{\hat{\tau}_{i-1}+1},\ldots,Z_{\tau(i)}\}$ and $\mathbf{Y}_{\tau(i)}(\kappa(i)) = \{Z_{\tau(i)+1},\ldots,Z_{\kappa(i)}\}$, $\hat{\tau}_{i-1} + 1 \leq \tau(i) < \kappa(i) \leq \hat{\tau}_i$, and
\begin{displaymath}
    \hat{\mathcal{Q}}(\mathbf{X}_{m},\mathbf{Y}_{m}(\kappa);\alpha) 
    = \frac{mn}{m+n}\hat{\mathcal{E}}(\mathbf{X}_{n},\mathbf{Y}_{m}(\kappa);\alpha).
\end{displaymath}
Among the $k$ candidates, the algorithm picks $\hat{\tau}_k = \hat{\tau}(i^*)$ where $i^* = \mathop{\rm argmax}_{i\in\{1,\ldots,k\}}\hat{\mathcal{Q}}(\mathbf{X}_{\hat{\tau}(i)},\mathbf{Y}_{\hat{\tau}(i)}(\hat{\kappa}(i));\alpha)$, and the corresponding test statistics is
\begin{equation}\label{e:test_statistics}
    \hat{q}_k = \hat{\mathcal{Q}}(\mathbf{X}_{\hat{\tau}_k},\mathbf{Y}_{\hat{\tau}_k(\hat{\kappa}_k)};\alpha).
\end{equation}
}

\textcolor{black}{The hierarchical significance testing performs a permutation test as follows. First, under the null hypothesis of no additional change points, the the observations within $\hat{\mathcal{C}}_i$, $i=1,\ldots,k$, are permuted to construct a new sequence of $\{Z_t\}_{t=1,\ldots,T}$. Then, obtain the test statistic $\hat{q}_k^{(r)}$ as \eqref{e:test_statistics} for each permutation. Repeat this $R$ times. Finally, the $k$th change point is identified as the new change point conditioning on $\{\hat{\tau}_1,\ldots,\hat{\tau}_{k-1}\}$ if 
\begin{displaymath}
    \frac{\# \{\hat{q}_k^{(r)} \geq \hat{q}_k \}}{R+1} \leq p_0.
\end{displaymath}
As a default, $p_0=0.05$ and $R=499$ are used.}

\textcolor{black}{Note that, like other CPD algorithms, change points should be located away from the boundaries of segments. This implies that the minimum time between detectable change points (i.e., the detection lower bound; see Definition 2.1 in \cite{cho2022two}) depends on the length of the segmented observations, as it follows a binary segmentation approach. The minimum allowable length is set to 30, which also serves as the minimum distance between change points. However, this length should ideally be extended if the empirical divergence measure \eqref{e:u-statistics} is small (see the discussion in Section 3.1 of\cite{matteson2014nonparametric}). Furthermore, according to Theorem 2 in the paper, any finite number of change points can be identified given a sufficiently large sample size. However, as the overall time between detectable change points decreases, the false positive rate increases.}

\newpage
\section{Illustration of Another Patient}
\label{sap:illustration}
\vspace*{12pt}

\begin{figure}[h]
\centering
\includegraphics[width=1\columnwidth,height=0.7\textwidth]{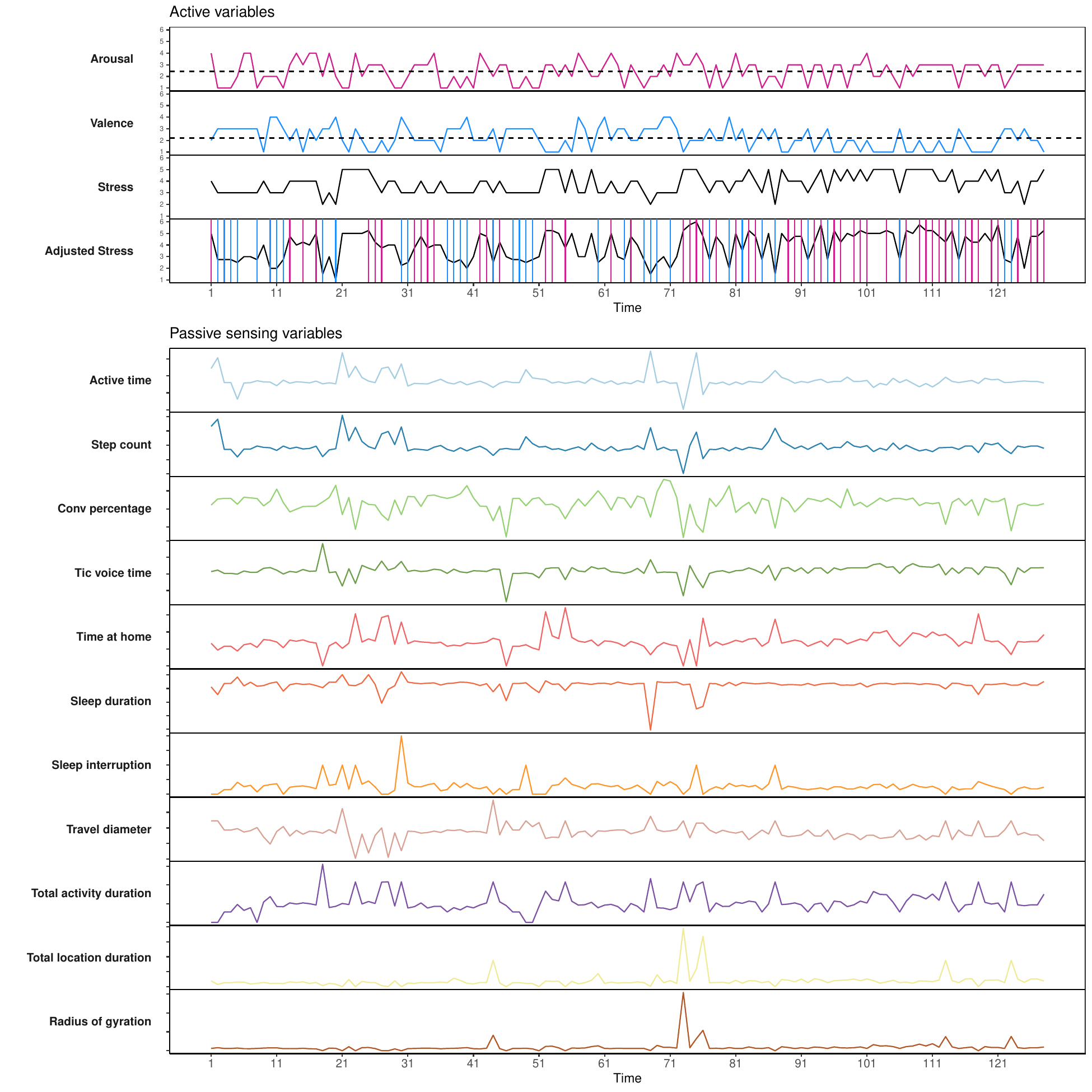}
\caption{Time plots of the stress level, arousal, valence, and adjusted stress level (top panel) and the 11 passive sensing variables for the patient \textcolor{black}{sternrelief9} (bottom panel). The description of the variable used in the plots can be found in Table \ref{tab:data}.}
\label{fig:initialization2}
\end{figure}

\begin{figure}[th]
\centering
\includegraphics[width=1\columnwidth,height=1\textwidth]{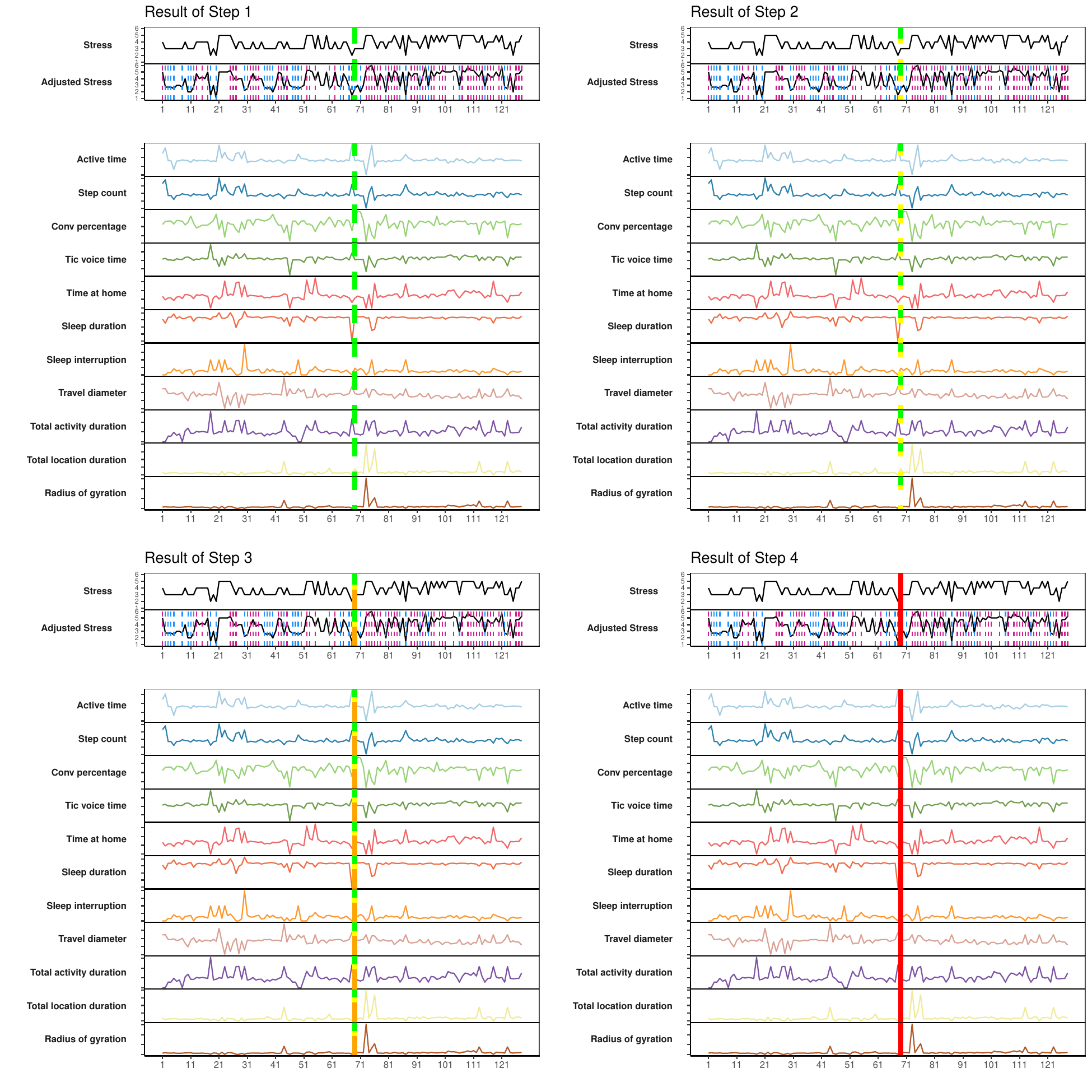}
\caption{Illustration of the results for the patient \textcolor{black}{sternrelief9 after each step. The top left panel represents the result after Step 1. The horizontal green dot-dashed line indicates the one initial change points at $t^*=69$. The top right panel shows the result after Step 2, where the validated time point overlaps the result in Step 1 and is marked as the thick yellow dotted line. The bottom two panels indicate the result after Steps 3 and 4, which are marked as the orange dashed and the red solid lines, respectively.}}
\label{fig:result_illustration2}
\end{figure}

\clearpage
\section{Simulation result}
\label{sap:simulation}
% Please add the following required packages to your document preamble:
% \usepackage{graphicx}
\begin{table}[h]
\centering
\resizebox{0.75\textwidth}{!}{%
\begin{tabular}{cccccccccccccccc}
\hline              & \multicolumn{15}{c}{Patient sternrelief9}                                                             \\ \hline
Step size     & \multicolumn{7}{c}{$s=2$ (50 iterations)}                  &  & \multicolumn{7}{c}{$s=5$ (23 iterations)}                  \\ \cline{1-8} \cline{10-16} 
Methods       & Reg & SVM & LDA & RF & Boost & Resp & Pred &  & Reg & SVM & LDA & RF & Boost & Resp & Pred \\ \cline{1-8} \cline{10-16} 
Rec           & 0.55 & 0.45 & 0.52 & \textbf{0.60} & 0.59 & 0.53 & 0.55 &  & 0.54 & 0.42 & 0.55 & 0.63 & \textbf{0.65} & 0.54 &  0.49  \\ \cline{1-1}
Prec          & 0.55 & 0.39 & 0.52 & \textbf{0.60} & 0.59 & 0.53 & 0.55 &  & 0.54 & 0.33 & 0.55 & 0.63 & \textbf{0.65} & 0.54 & 0.49 \\ \cline{1-1}
Acc           & 0.52 & 0.53 & 0.54 & \textbf{0.62} & 0.60 & 0.52 & 0.54  &  & 0.53 & 0.43 & 0.56 & 0.63 & \textbf{0.65} & 0.54 & 0.50 \\ \cline{1-1}
$\mbox{F1}_1$ & 0.51 & \textbf{0.68} & 0.61 & \textbf{0.68} & 0.66 & 0.56 & 0.58 &  & 0.48 & 0.59 & 0.60 & 0.67 & \textbf{0.68} & 0.58 & 0.52 \\ \cline{1-1}
$\mbox{F1}_2$ & \textbf{0.53} & 0.09 & 0.43 & 0.52 & 0.51 & 0.48 & 0.50 &  & 0.57 & 0.08 & 0.50 & 0.60 & \textbf{0.62} & 0.50 & 0.47  \\ \hline
                & \multicolumn{15}{c}{Patient sternrelief22}                                                             \\ \hline
Step size     & \multicolumn{7}{c}{$s=2$ (31 iterations)}                  &  & \multicolumn{7}{c}{$s=5$ (18 iterations)}                  \\ \cline{1-8} \cline{10-16} 
Methods       & Reg & SVM & LDA & RF & Boost & Resp & Pred &  & Reg & SVM & LDA & RF & Boost & Resp & Pred \\ \cline{1-8} \cline{10-16} 
Rec           & 0.91 & 0.91 & 0.91 & \textbf{0.93} & 0.90 & 0.55 & 0.54  &  & 0.94 & 0.94 & 0.93 & \textbf{0.96} & 0.94 & 0.59 & 0.63  \\ \cline{1-1}
Prec          & 0.87 & 0.86 & 0.88 & \textbf{0.89} & 0.88 & 0.55 & 0.53  &  & 0.86 & 0.86 & 0.87 & 0.88 & \textbf{0.89} & 0.56 & 0.59 \\ \cline{1-1}
Acc           & 0.88 & 0.88 & 0.90 & \textbf{0.91} & 0.90 & 0.57 & 0.55  &  & 0.91 & 0.91 & 0.92 & \textbf{0.93} & \textbf{0.93} & 0.52 & 0.57 \\ \cline{1-1}
$\mbox{F1}_1$ & 0.91 & 0.90 & 0.92 & \textbf{0.93} & 0.92 & 0.66 & 0.64 &  & 0.94 & 0.94 & 0.95 & \textbf{0.96} & \textbf{0.96} & 0.61 & 0.65 \\ \cline{1-1}
$\mbox{F1}_2$ & 0.85 & 0.84 & 0.85 & \textbf{0.87} & 0.85 & 0.43 & 0.41 &  & 0.83 & 0.83 & 0.84 & \textbf{0.87} & 0.86 & 0.39 & 0.43  \\ \hline
                & \multicolumn{15}{c}{Patient wynneprotect23}                                                             \\ \hline
Step size     & \multicolumn{7}{c}{$s=2$ (26 iterations)}                  &  & \multicolumn{7}{c}{$s=5$ (10 iterations)}                  \\ \cline{1-8} \cline{10-16} 
Methods       & Reg & SVM & LDA & RF & Boost & Resp & Pred &  & Reg & SVM & LDA & RF & Boost & Resp & Pred \\ \cline{1-8} \cline{10-16} 
Rec           & 0.56 & \textbf{0.67} & 0.66 & 0.63 & 0.56 & 0.61 & 0.53  &  & 0.58 & \textbf{0.68} & \textbf{0.68} & 0.60 & 0.64 & 0.52 & 0.50  \\ \cline{1-1}
Prec          & 0.53 & \textbf{0.65} & 0.63 & 0.59 & 0.54 & 0.57 & 0.52  &  & 0.55 & \textbf{0.66} & \textbf{0.66} & 0.57 & 0.59 & 0.52 & 0.50 \\ \cline{1-1}
Acc           & 0.58 & \textbf{0.77} & 0.75 & 0.68 & 0.62 & 0.62 & 0.53  &  & 0.62 & \textbf{0.78} & \textbf{0.78} & 0.66 & 0.66 & 0.48 & 0.44 \\ \cline{1-1}
$\mbox{F1}_1$ & 0.69 & \textbf{0.85} & 0.84 & 0.78 & 0.73 & 0.73 & 0.65 &  & 0.73 & \textbf{0.86} & \textbf{0.86} & 0.77 & 0.76 & 0.58 & 0.53 \\ \cline{1-1}
$\mbox{F1}_2$ & 0.32 & \textbf{0.46} & 0.44 & 0.40 & 0.32 & 0.38 & 0.30 &  & 0.34 & \textbf{0.48} & \textbf{0.48} & 0.37 & 0.41 & 0.32 & 0.30  \\ \hline \hline        
                & \multicolumn{15}{c}{Patient gayevskyred18}                                                            \\ \hline
Step size     & \multicolumn{7}{c}{$s=2$ (50 iterations)}                  &  & \multicolumn{7}{c}{$s=5$ (24 iterations)}                  \\ \cline{1-8} \cline{10-16} 
Methods       & Reg & SVM & LDA & RF & Boost & Resp & Pred &  & Reg & SVM & LDA & RF & Boost & Resp & Pred \\ \cline{1-8} \cline{10-16} 
Rec           & 0.01 & 0.00 & 0.45 & 0.45 & \textbf{0.49} & 0.27& 0.20 &  & 0.00 & 0.00 & 0.47 & 0.44  & \textbf{0.50} & 0.29 & 0.32   \\ \cline{1-1}
Prec          & 0.33 & 0.00 & 0.46 & 0.43 & \textbf{0.47} & 0.25 & 0.20 &  & 0.00 & 0.00 & 0.49 & 0.45 & \textbf{0.50} & 0.30 & 0.33  \\ \cline{1-1}
Acc           & 0.01 & 0.00 & 0.46 & 0.44 & \textbf{0.49} & 0.26 & 0.21 &  & 0.00 & 0.00 & 0.48 & 0.45 & \textbf{0.51} & 0.30 & 0.32 \\ \cline{1-1}
$\mbox{F1}_1$ & 0.00 & 0.00 & \textbf{0.46} & 0.44 & 0.44 & 0.32 & 0.26 &  & 0.00 & 0.00 & 0.46 & 0.45 &  \textbf{0.50} & 0.36 & 0.38 \\ \cline{1-1}
$\mbox{F1}_2$ & 0.04 & 0.00 & 0.53 & 0.51 & \textbf{0.62} & 0.21 & 0.23  &  & 0.00 & 0.00 & 0.56 & 0.52  & \textbf{0.59} & 0.28 & 0.28 \\  \cline{1-1}
$\mbox{F1}_3$ & 0.00 & 0.00 & 0.32 & 0.34 & \textbf{0.35} & 0.21 & 0.10 &  & 0.00 & 0.00 & 0.41 & 0.37 & \textbf{0.42} & 0.24 & 0.31 \\ \hline
              & \multicolumn{15}{c}{Patient sternrelief4}                                                            \\ \hline
Step size     & \multicolumn{7}{c}{$s=2$ (41 iterations)}                  &  & \multicolumn{7}{c}{$s=5$ (18 iterations)}                  \\ \cline{1-8} \cline{10-16} 
Methods       & Reg & SVM & LDA & RF & Boost & Resp & Pred &  & Reg & SVM & LDA & RF & Boost & Resp & Pred \\ \cline{1-8} \cline{10-16} 
Rec           & 0.25 & 0.24 & 0.52 & \textbf{0.56} & \textbf{0.56} & 0.43 & 0.39  &  & 0.30 & 0.29 & \textbf{0.52} & \textbf{0.52} & 0.51 & 0.48 & 0.40  \\ \cline{1-1}
Prec          & 0.39 & 0.43 & \textbf{0.61} & 0.58 & 0.57 & 0.50 & 0.47 &  & 0.39 & 0.43 & 0.60 & 0.57 & 0.54 & \textbf{0.69} & 0.32  \\ \cline{1-1}
Acc           & 0.25 & 0.23 & 0.43 & \textbf{0.45} & 0.44 & 0.44 & \textbf{0.45} &  & 0.29 & 0.27 & 0.42 & 0.43 & 0.41 & \textbf{0.56} & 0.47 \\ \cline{1-1}
$\mbox{F1}_1$ & 0.21 & 0.12 & 0.23 & 0.28 & 0.24 & 0.52 & \textbf{0.58} &  & 0.24 & 0.15 & 0.27 & 0.32 & 0.26 & \textbf{0.63} & 0.52 \\ \cline{1-1}
$\mbox{F1}_2$ & 0.41 & 0.41 & \textbf{0.50} & \textbf{0.50} & 0.48 & 0.31 & 0.34  &  & 0.44 & 0.44 & 0.48 & 0.47 & 0.47 & \textbf{0.52} & 0.51 \\ \cline{1-1}
$\mbox{F1}_3$ & 0.00 & 0.00 & 0.57 & \textbf{0.68} & \textbf{0.68} & 0.52 & 0.26 &  & 0.00 & 0.00 & 0.55 &  \textbf{0.57} & 0.55 & 0.24 & 0.00 \\ \hline
              & \multicolumn{15}{c}{Patient sternrelief5}                                                            \\ \hline
Step size     & \multicolumn{7}{c}{$s=2$ (50 iterations)}                  &  & \multicolumn{7}{c}{$s=5$ (30 iterations)}                  \\ \cline{1-8} \cline{10-16} 
Methods       & Reg & SVM & LDA & RF & Boost & Resp & Pred &  & Reg & SVM & LDA & RF & Boost & Resp & Pred \\ \cline{1-8} \cline{10-16} 
Rec           & 0.03 & 0.04 & 0.57 & \textbf{0.58} & 0.57 & 0.28 & 0.31  &  & 0.02 & 0.02 & 0.59 & 0.59 & \textbf{0.60} & 0.33 & 0.33 \\ \cline{1-1}
Prec          & 0.33 & 0.33 & 0.62 & \textbf{0.64} & 0.59 & 0.30 & 0.33 &  & 0.33 & 0.33 & \textbf{0.55} & \textbf{0.55} & 0.54 & 0.36 & 0.34 \\ \cline{1-1}
Acc           & 0.05 & 0.06 & \textbf{0.49} & 0.48 & 0.47 & 0.32 & 0.38 &  & 0.03 & 0.03 & \textbf{0.55} & 0.53 & 0.54 & 0.33 & 0.34 \\ \cline{1-1}
$\mbox{F1}_1$ & 0.00 & 0.00 & 0.50 & 0.51 & \textbf{0.52} & 0.20 & 0.19 &  & 0.00 & 0.00 & 0.49 & 0.51 & \textbf{0.52} & 0.18 & 0.16 \\ \cline{1-1}
$\mbox{F1}_2$ & 0.17 & 0.21 & 0.42 & 0.39 & 0.37 & 0.44 & \textbf{0.52}  &  & 0.14 & 0.14 & 0.29 & 0.26 & 0.26 & 0.43 & \textbf{0.47} \\ \cline{1-1}
$\mbox{F1}_3$ & 0.00 & 0.00 & \textbf{0.64} & \textbf{0.64} & 0.60 & 0.21 & 0.23 &  & 0.00 & 0.00 & \textbf{0.80} & 0.78 & 0.78 & 0.34 & 0.31 \\ \hline
\end{tabular}%
}
\caption{Performance measures for classification methods and benchmarks for the six patients. The first three patients have $N^*=2$ final segments and the last three patients have $N^*=3$ final segments. The best performance in each measure among prediction methods is marked in bold (more than one number in each row is bold if they are tied).}
\label{tab:result}
\end{table}

\end{document}